# Detection of the anomalous velocity with sub-picosecond time resolution in semiconductor nanostructures

*Shekhar Priyadarshi,[*] Klaus Pierz, and Mark Bieler[†]*

Physikalisch-Technische Bundesanstalt, 38116 Braunschweig, Germany

We report on the time-resolved detection of the anomalous velocity, constituting charge carriers moving perpendicular to an electric driving field, in undoped GaAs quantum wells. For this we optically excite the quantum wells with circularly polarized femtosecond laser pulses, thereby creating a state which breaks time-inversion symmetry. We then employ a quasi single cycle terahertz pulse as electric driving field to induce the anomalous velocity. The electromagnetic radiation emitted from the anomalous velocity is studied with a sub-picosecond time resolution and reveals intriguing results. We are able to distinguish between intrinsic (linked to the Berry curvature) and extrinsic (linked to scattering) contributions to the anomalous velocity both originating from the valence band and observe local energy space dependence of the anomalous velocity. Our results thus constitute a significant step towards non-invasive probing of the anomalous velocity locally in the full energy/momentum space and enable the investigation of many popular physical effects such as anomalous Hall effect and spin Hall effect on ultrafast time scales.

PACS numbers: 82.53.Mj, 72.25.Dc, 78.47.J-, 71.35.-y, 73.63.Hs

---

[*] e-mail: shekhar.priyadarshi@ptb.de
[†] e-mail: mark.bieler@ptb.de

The movement of charge carriers perpendicular to an electric driving field – even without any magnetic field – constitutes one of the most intriguing properties of carriers in solids. This anomalous movement results in an observable velocity [1] referred to as anomalous velocity $\langle v \rangle_a$ [2–4]. It is at the origin of fascinating physical phenomena [3,5–10], with the spin Hall [5,6] and anomalous Hall [7,9,11] effects being prominent examples, and has direct impact on spintronics [12,13], topological insulators [7,8,14–16], and novel quantum computers [15,17,18]. It is well known that different microscopic effects contribute to $\langle v \rangle_a$ and these are typically divided into intrinsic and extrinsic contributions linked to the Berry curvature (**Ω**) [2,3,19] and carrier scattering processes [11,20], respectively. However, despite numerous previous studies no straightforward way exists to distinguish between extrinsic and intrinsic contributions [21]. Moreover, being equally important $\langle v \rangle_a$ has not been investigated in the ultrafast regime in which coherence and relaxation processes might have a profound influence on its dynamics.

In this letter, we report on the detection of $\langle v \rangle_a$ in undoped GaAs quantum wells (QW) with sub-picosecond time resolution and show that such experiments are an ideal tool to distinguish between extrinsic and intrinsic effects. For this purpose free-space, single-cycle terahertz (THz) pulses are used to accelerate carriers being excited by femtosecond optical pulses in the QWs. This combined optical and THz excitation induces $\langle v \rangle_a$ [3,22–24] on an ultrafast time scale, which, in turn emits THz radiation into free space. We perform a time-resolved study of this THz emission versus delay between the optical and THz excitation pulses. Our measurements confirm that $\langle v \rangle_a$ results almost entirely from the valence band but also yield two other striking results: (i) The strength of $\langle v \rangle_a$ is significantly influenced by energy relaxation in addition to spin relaxation. (ii) Extrinsic effects dominate over intrinsic effects, although we also detect contributions from **Ω**.

We first introduce extrinsic and intrinsic contributions to $\langle v \rangle_a$ in more detail. The intrinsic contribution $\langle v \rangle_{a,\text{int}}$ is due to **Ω** [25] and has first been theoretically studied in [1,2,26–28]. The resulting microscopic velocities are proportional to $\mathbf{E} \times \mathbf{\Omega}$, with **E** being the electric bias field, see Fig. 1a [3,19]. The extrinsic contribution $\langle v \rangle_{a,\text{ext}}$ appears due to scattering of carriers with impurities or phonons [11,20,29] and can be divided into three components [29]. The first two components result from coordinate shifts of carriers during scattering events involving symmetric scattering rates. In the first case, such scattering processes induce a microscopic velocity known as side-jump velocity. Upon modification of the momentum-space carrier distribution by the electrical bias the macroscopic side-jump velocity $\langle v \rangle_{a,ext,sj}$ results, see Fig. 1b [11,29]. In the second case carriers lose or gain energy during scattering-induced coordinate shifts along the electrical bias which simultaneously push carriers in momentum space in a direction transverse to the electrical bias. This component is called anomalous carrier distribution and is temporally indistinguishable from $\langle v \rangle_{a,ext,sj}$. Therefore, in the following discussion we do not distinguish between the anomalous distribution and $\langle v \rangle_{a,ext,sj}$. The third extrinsic component, referred to as skew scattering $\langle v \rangle_{a,ext,sk}$, is linked to asymmetric scattering of carriers in momentum space. This effect transfers the electrical-bias-induced asymmetry of the carrier distribution to the direction transverse to the electrical bias, see Fig. 1c.

As samples we have chosen (110)- and (001)-oriented, undoped, nominally-symmetric GaAs/Al$_{0.3}$Ga$_{0.7}$As QWs, mainly because their bandstructure and most relevant relaxation times are well known and they are available in a high quality. Moreover, the different orientations exhibit different spin relaxation times which will prove helpful for the data analysis. The (110)-oriented samples have well widths of 8 nm, 15 nm, 18 nm, 24 nm, and 28 nm with the crystallographic directions [001], [1$\bar{1}$0], and [110] taken as the $\hat{x}$, $\hat{y}$, and $\hat{z}$ directions, respectively [30]. The (001)-oriented sample has a well width of 28 nm with the crystallographic directions [1$\bar{1}$0], [110], and [001] taken as the $\hat{x}$, $\hat{y}$, and $\hat{z}$ directions, respectively.

For the generation and detection of $\langle v \rangle_a$ we adopt a recently proposed scheme for probing of $\mathbf{\Omega}$ in semiconductor nanostructures [3] but additionally consider extrinsic effects in our data analysis. The output of a femtosecond laser (150 fs long optical pulses whose centre frequency can be tuned between 1.51 eV and 1.55 eV to excite heavy-hole excitons in our samples) is split into three beams. The first beam (referred to as optical pump beam) with a pulse energy of 2 nJ is circularly polarized and focused to a spot size of ~200 μm exciting a carrier density of ~$10^{11}$cm$^{-2}$ in the QWs. This excitation creates a state which breaks time-inversion symmetry [5], being a prerequisite for the observation of $\langle v \rangle_a$ in GaAs. The second beam is linearly polarized and used to generate THz pulses $E_{THz}$ via optical rectification in a 2 mm thick (110)-oriented zinc telluride (ZnTe) crystal. These THz pulses act as time-dependent electric bias. They are focused at normal incidence on the same position of the samples as the optical pump beam yielding a peak electric field inside the sample of ~10 Vcm$^{-1}$. The polarization of $E_{THz}$ is parallel to the $\hat{x}$ axis of the samples as ensured by the orientation of the ZnTe crystal and an additional THz polarizer. The combined optical/THz excitation induces $\langle v \rangle_a$ in the plane of the QWs along the $\hat{y}$ direction. Being induced on an ultrafast time scale, $\langle v \rangle_a$ emits THz radiation, whose electric field is denoted as $\langle \dot{v} \rangle_a$ since it is proportional to the time-derivative of $\langle v \rangle_a$ and perpendicularly polarized compared to $E_{THz}$. The third optical beam, referred to as probe beam, is linearly polarized and used to detect $\langle \dot{v} \rangle_a$ employing electro-optic sampling in a 1 mm thick (110)-oriented ZnTe crystal [31]. In order to suppress $E_{THz}$ and only detect $\langle \dot{v} \rangle_a$ the detection ZnTe crystal is appropriately oriented and two THz polarizers are used. All experiments are performed at 10 K except as explicitly mentioned.

For demonstration of the existence of $\langle v \rangle_a$ we start with experiments on 28 nm (110)-oriented GaAs QWs. In Fig. 2a we plot the amplitude of $\langle \dot{v} \rangle_a$ versus delay between the optical and $E_{THz}$ excitation pulses for different helicities of the optical excitation. The helicity of circular polarization of the optical pump determines the spin polarization of carriers [32–34]. Reversing the helicity of the optical pump reverses the spin polarization and, consequently, reverses $\langle v \rangle_a$. Moreover the amplitude of $\langle v \rangle_a$ crucially depends on the delay between the optical and $E_{THz}$ excitation. A slight difference between the two traces shown in Fig. 2a appears because of the imperfect suppression of THz signals polarized along the x-direction [35]. Such signals appear due to THz induced transport of optically excited carriers along the x-direction being independent of the helicity of the optical pump beam. Therefore, for the remainder of this paper we always subtract signals resulting from left-handed and right-handed circularly polarized pump to obtain a background-free signal. The difference between the two curves of Fig. 2a is shown in Fig. 2b. The amplitude of this difference signal is reduced if the polarization of the optical pump beam is changed from circular to elliptical polarization (data not shown), proving that our signal only appears due to circular polarization of the optical pump beam. Again, in Fig. 2a a bi-exponential decay of the curve is obtained with fast and slow decay rates of ~(3 ps)$^{-1}$ and ~(10 ps)$^{-1}$, respectively. Repeating the experiments on 15 nm (110)-oriented QWs we obtain decay rates of ~(6 ps)$^{-1}$ and ~(90 ps)$^{-1}$. We expect that the slow decay rate corresponds to spin relaxation diminishing $\langle v \rangle_a$, since this process destroys spin polarization of carriers and establishes time-inversion symmetry. Therefore, the measured signal has to result from the valence band with previously obtained spin relaxation rates $w^{sf} \approx (50\text{ ps})^{-1}$ at 4 K [36,37]. We can rule out that $\langle v \rangle_a$ originates from the conduction band, with typical spin relaxation rates of $(300\text{ ps})^{-1}$ at 10 K [38]. Two other observations confirm this conclusion. First, only a single exponential decay with a rate of $\sim(1\text{ ps})^{-1}$ is obtained upon increasing the temperature to 100 K, see Fig. 2d. In contrast, the spin relaxation time in the conduction band of (110)-oriented samples will increase with increasing temperature [38]. Second, the decay rate at 80 K increases with quantum well width, see Fig. 2e. This can be explained with spin relaxation being enhanced due to bandmixing in the valence band which is more pronounced in wider QWs [36].

Having linked the slow decay rate to spin relaxation in the valence band we will now explore the origin of the fast decay rate of the delay scans. In Fig. 2c we plot the same delay scan as shown in Fig. 2b but measured on (001)-oriented 28 nm GaAs QWs. A very similar dependence with decay rates of $\sim(2\text{ ps})^{-1}$ and $\sim(9\text{ ps})^{-1}$ is obtained as compared to the (110) orientation. Therefore, we rule out that momentum relaxation of an optically-induced asymmetric hole distribution, which is present for the (110) orientation but not for the (001) orientation [33,39], is at the main origin of the fast decay rate. Instead we believe that energy relaxation of holes predominantly causes the fast decay, see also supplementary material. This is plausible since $\langle v \rangle_a$ is dependent on carrier energy [3,11,14,40].

So far we have only established the existence of $\langle v \rangle_a$ without being able to distinguish between extrinsic and intrinsic contributions. The existence of $\langle v \rangle_{a,ext}$ is indicated by the dependence of the amplitude of $\langle v \rangle_a$ on temperature (Fig. 1d), since $\mathbf{\Omega}$ is not expected to increase with temperature. Yet as shown below a more straightforward way to distinguish between $\langle v \rangle_{a,int}$ and $\langle v \rangle_{a,ext}$ is to analyze the shape of $\langle \dot{v} \rangle_a$. For this purpose we perform delay scans such as the ones plotted in Fig. 2b at different temporal instances $t$ of $\langle \dot{v} \rangle_a(t)$. This experiment leads to contour plots for the (001)- and (110)-oriented 28 nm QWs, see Fig. 3a and c, respectively. Here, the x and y axes denote the time instance of $\langle \dot{v} \rangle_a$ and the delay between optical and THz excitations, respectively. Comparing the contour plots obtained from the (110)- and (001)-oriented samples with each other two main features are observed. For the (110) orientation $\langle \dot{v} \rangle_a$ is phase shifted to $E_{THz}(t)$ at early delay times but directly follows $E_{THz}(t)$ at later delay times, see Fig. 3d. This initial phase shift, which decays with the fast decay rate observed in Fig. 2b, is considerably reduced in the (001)-oriented GaAs QWs, see Fig. 3b.

To understand these differences we have to discuss the expected temporal shape of the different intrinsic and extrinsic contributions to $\langle v \rangle_a$. We give a simple intuitive explanation below. A more detailed discussion with the same qualitative results can be found in the supplementary material [41]. As introduced earlier $\langle v \rangle_{a,int}$ is proportional to the microscopic velocity given by $\mathbf{E}_{THz} \times \mathbf{\Omega}$, and hence its time dependence can be expressed as $\langle v \rangle_{a,int} \propto E_{THz}(t)$ [3]. Unlike $\langle v \rangle_{a,int}$, which does not require any modification of the carrier distribution by $E_{THz}$, $\langle v \rangle_{a,ext,sj}$ requires the carrier distribution to be modified by $E_{THz}$, see Fig. 1b. This modified carrier distribution $\rho$ relaxes to equilibrium due to scattering with impurities or phonons. The corresponding dynamic equation is given by $\dot{\rho} + w_{\mathbf{k}}^s \rho \propto E_{THz}$, with $w_{\mathbf{k}}^s$ being the symmetric scattering rate and ignoring the much smaller asymmetric scattering rate $w_{\mathbf{k}}^a$ [20]. This yields $\langle v \rangle_{a,ext,sj} \propto \rho \propto \{E_{THz} * [u(t) \exp(-w_{\mathbf{k}}^s t)]\}$, with $u(t)$ being the step function and $*$ denoting convolution. Skew scattering is linked to asymmetric scattering for which the carriers are scattered preferably in a certain direction. Consequently, the whole distribution will be shifted along this direction. However, the equilibrium distribution does not contribute to $\langle v \rangle_{a,ext,sk}$, only the non-equilibrium carrier distribution $\rho$ which is already modified by $E_{THz}$ will be rotated at right angle due to skew scattering, see Fig. 1c. The dynamics of the skew-scattered carrier distribution $\rho_{sk}$ is given by the equation $\dot{\rho}_{sk} + w_{\mathbf{k}}^s \rho_{sk} = w_{\mathbf{k}}^a \rho$. This yields $\langle v \rangle_{a,ext,sk} \propto \rho_{sk} \propto w_{\mathbf{k}}^a \{E_{THz} * [u(t) \, t \, \exp(-w_{\mathbf{k}}^s t)]\}$. In our simple model given here and in [41], we ignore Coulomb interaction. Coulomb interaction between electrons (or holes) with opposite spins is responsible for the Coulomb drag effect [20]. However, this effect is not important in our case since we excite spin-polarized carriers.

Our simple analysis clearly shows that the shape of $\langle v \rangle_a$ and, therefore, also of $\langle \dot{v} \rangle_a$ will differ significantly for their various components. For symmetric scattering rates lying in the picosecond range, $\langle \dot{v} \rangle_{a,int}$ and $\langle \dot{v} \rangle_{a,ext,sk}$ peak at times where $E_{THz}(t)$ is zero, while $\langle \dot{v} \rangle_{a,ext,sj}$ almost directly follows $E_{THz}$, see last row of Fig. 1. Comparing these expected temporal shapes of $\langle \dot{v} \rangle_a$ with the measured shapes of Fig. 3b and d, a striking conclusion can be made. In (001)-oriented samples the main contribution to $\langle v \rangle_a$ arises from $\langle v \rangle_{a,ext,sj}$ and not from $\langle v \rangle_{a,int}$ as previously assumed in Refs. [3,42–46]. Only in (110)-oriented samples $\langle v \rangle_{a,int}$ and/or $\langle v \rangle_{a,ext,sk}$ seem to significantly contribute to $\langle v \rangle_a$, especially at early time delays. Before discussing these contributions we like to point out the influence of energy relaxation on $\langle v \rangle_{a,ext,sj}$. In the (001)-oriented sample where the phase shift of $\langle \dot{v} \rangle_a$ as compared to $E_{THz}$ is small the biexponential decay of the delay scan (Fig. 2c) has to be caused by $\langle v \rangle_{a,ext,sj}$. Thus, $\langle v \rangle_{a,ext,sj}$ not only depends on spin polarization but also on energy and momentum [11]. The fact that this dependence becomes clearly visible in our results confirms the strength of our experimental technique and paves the way for local (energy space) probing of $\langle v \rangle_a$.

We will now argue that the initial phase-shift of $\langle \dot{v} \rangle_a$ in the (110)-oriented samples results from $\langle v \rangle_{a,int}$. In principle, both $\langle v \rangle_{a,int}$ and $\langle v \rangle_{a,ext,sk}$ will lead to a phase shift of $\langle \dot{v} \rangle_a$ as compared to $E_{THz}$. However, the contribution from $\langle v \rangle_{a,ext,sk}$ will be considerably slower than the contribution from $\langle v \rangle_{a,int}$, see Fig. 1, and, thus, lead to stretched THz traces. We do not observe THz traces which are stretched as compared to $E_{THz}$, but rather compressed in time, see Fig. 3d. This observations can only be explained by considering $\langle v \rangle_{a,int}$, proving the existence of this component. Since $\mathbf{\Omega}$ of the first heavy-hole state is zero at $\mathbf{k} = 0$ [3,40], $\langle v \rangle_{a,int}$ will predominantly vanish with energy relaxation and not with spin relaxation as $\langle v \rangle_{a,ext,sj}$. We believe that this is the reason why the phase shift of $\langle \dot{v} \rangle_a$ decreases with increasing delay. We can rule out that the biexponential decay of the delay scans is predominantly caused by biexponential energy relaxation [47]. In such a case the contour plot and individual THz traces shown in Figs. 3c and d will not show any delay-dependent phase shift.

Simulations based on above arguments considering $\langle v \rangle_{a,int}$ and $\langle v \rangle_{a,ext,sj}$ are plotted in Fig. 3e-h, see [41]. The simulations show very good resemblance with the experiments and provide ratios $\frac{\langle v \rangle_{a,int}}{\langle v \rangle_{a,ext,sj}} \sim 0.04$ and $\frac{\langle v \rangle_{a,int}}{\langle v \rangle_{a,ext,sj}} \sim 0.6$ for (001)-oriented and (110)-oriented GaAs QWs, respectively. Although we cannot fully rule out that also $\langle v \rangle_{a,ext,sk}$ contributes to $\langle v \rangle_a$, all features of the experimental results can be explained without considering $\langle v \rangle_{a,ext,sk}$. The stronger intrinsic contribution to $\langle v \rangle_a$ in the (110)-oriented samples compared to (001)-oriented samples can be attributed to bandmixing in the valence band [3,40]. Bandmixing is known to enhance $\mathbf{\Omega}$ and it is stronger in (110)-oriented than in (001)-oriented GaAs QWs [39]. Bandmixing also enhances the overall anomalous velocity response which is apparent from the increasing amplitude of $\langle \dot{v} \rangle_a$ with increasing quantum well width of the (110)-oriented samples, see Fig. 1e [3,40]. We have also measured similar behavior in (001)-oriented GaAs QWs.

Finally we comment on small oscillations, which are superimposed on the delay scans at very early time delays, see Fig. 3a and c. We believe that the oscillations with a frequency of ~3 THz appear because of the presence of a shift current component [4] being $2^{nd}$ order in optical field and first order in THz field. We have ruled out that the oscillations are due to plasma, Rabi, or Zitterbewegung effects. In any case it should be emphasized that the oscillations vanish within ~0.6 ps while the phase shift of $\langle \dot{v} \rangle_a$ persists for several picoseconds. In fact, in (110)-oriented QWs we even observe a phase shift after 6 ps time delay, see Fig. 3(d) confirming that the observed phase shift is not an artifact appearing due to the aforementioned oscillations.

In conclusion, we have detected $\langle v \rangle_a$ showing a sub-picosecond time response in GaAs QWs. We not only identified $\langle v \rangle_{a,int}$ and $\langle v \rangle_{a,ext,sj}$ contributions but could also estimate their relative strengths. This study shows that even in a clean and cold semiconductor extrinsic effects may not be ignored and may even constitute the major contribution to $\langle v \rangle_a$. Our observations will have important implications on future work regarding Berry curvature and spin transports in semiconductors. Finally we note that the time-resolved detection of $\langle v \rangle_a$ as advocated in this work can be applied to a wide range of materials enabling direct probing of topological properties.

**Acknowledgement**

The authors thank E.O. Göbel for comments on an initial version of the manuscript and the Deutsche Forschungsgemeinschaft for financial support.

**Figure 1**

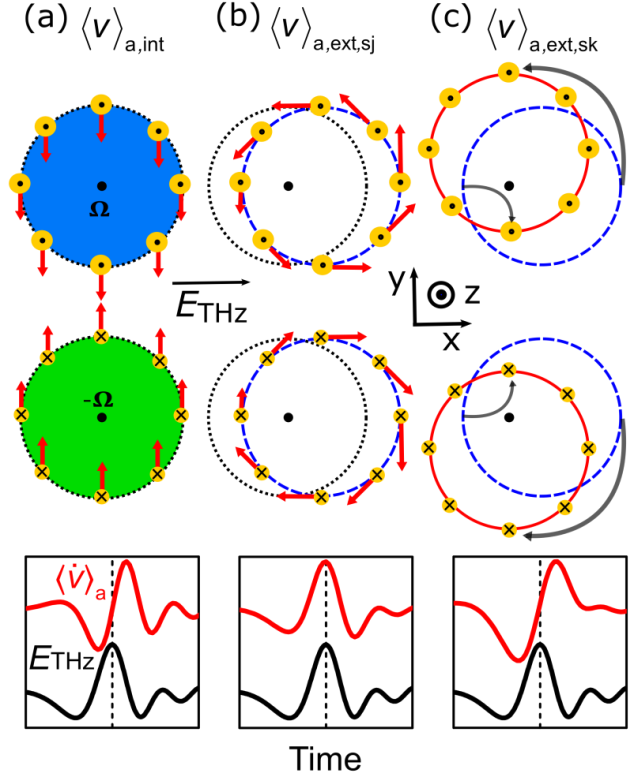

Fig. 1: (Color online) Extrinsic and intrinsic contributions. The upper two rows denote two different spin subbands. The third row shows simulated terahertz traces ($\langle \dot{v} \rangle_a$) emitted from the different components of $\langle v \rangle_a$ with respect to the time-dependent electric bias $E_{\text{THz}}$. (a) Intrinsic contribution $\langle v \rangle_{a,\text{int}}$ due to the Berry curvature $\mathbf{\Omega}$. The blue ($\mathbf{\Omega}$) and green ($-\mathbf{\Omega}$) colors show Berry curvatures pointing along opposite z directions. The red solid arrows indicate microscopic anomalous velocities (with the arrow length denoting the strength). The black dotted circles represent the energy of carriers being excited with circularly polarized light in one spin subband (without considering spin splitting). The big and small yellow balls indicate large and small density of carriers, respectively. The dots and crosses within the balls indicate spin along +z (spin up) and −z (spin down) directions, respectively. (b) Side-jump velocity $\langle v \rangle_{a,\text{ext,sj}}$ for the optically excited carrier distribution being displaced by $E_{\text{THz}}$. The blue dashed circles represent the energy of carriers modified by $E_{\text{THz}}$. (c) Skew scattering resulting in $\langle v \rangle_{a,\text{ext,sk}}$. The red solid curves represent the energy of carriers modified by scattering processes. The thick (thin) gray arrow denotes the scattering path for carriers from states with large (small) momentum along *x* to states with large (small) momentum along *y*. The normal microscopic velocities resulting from the band dispersion are not indicated.

**Figure 2**

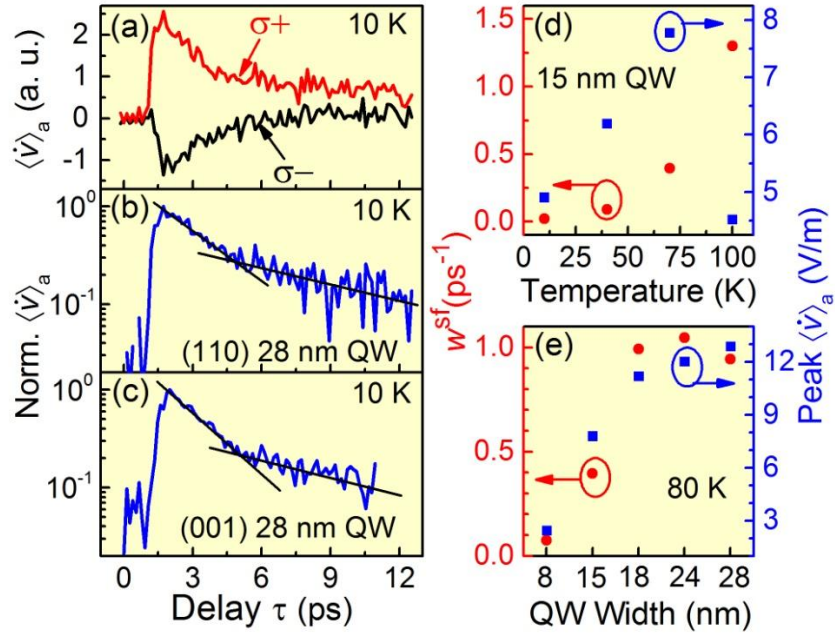

Fig. 2: (Color online) (a) Amplitude of the emitted terahertz radiation $\langle \dot{v} \rangle_a$ from (110)-oriented 28 nm GaAs QWs at 10 K versus delay between optical excitation and $E_{\text{THz}}$ for different optical helicities (σ+ and σ-). (b) Difference of the delay scans shown in (a), referred to as delay scan, plotted on a logarithmic scale showing a bi-exponential decay. (c) Delay scan measured in (001)-oriented 28 nm GaAs QWs and plotted on a logarithmic scale showing a bi-exponential decay. (d) Spin relaxation rates $w^{\text{sf}}$ and amplitude of $\langle \dot{v} \rangle_a$ versus temperature in (110)-oriented 15 nm GaAs QWs. (e) Spin relaxation rates $w^{\text{sf}}$ and amplitude of $\langle \dot{v} \rangle_a$ for different (110)-oriented QWs at 80 K.

**Figure 3**

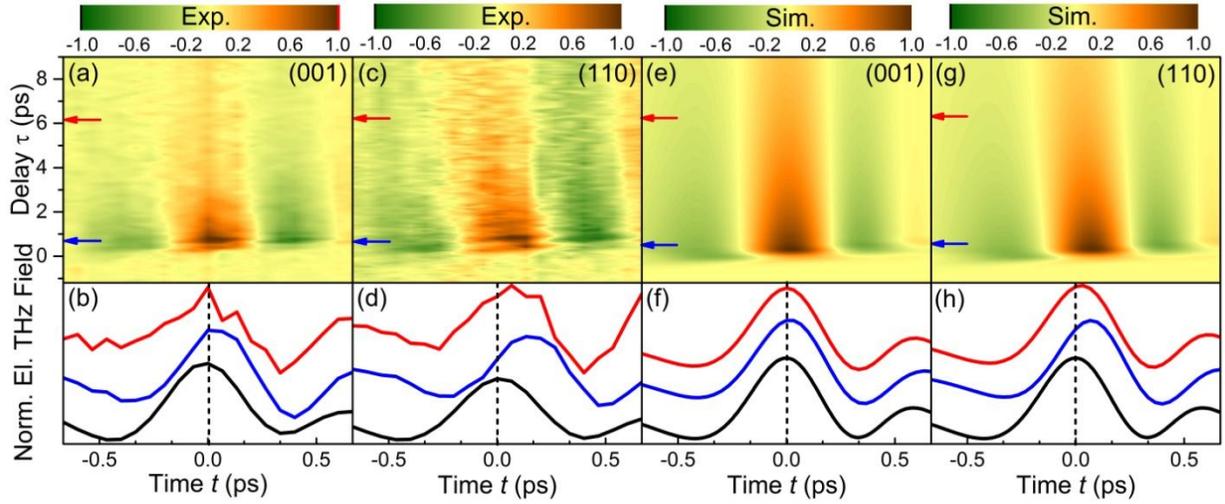

Fig. 3: Upper row: Shown are contour plots in which $\langle \dot{v} \rangle_a$ is plotted versus time (x axis) and delay between the optical and $E_{\mathrm{THz}}$ excitation pulses (y axis) in (001)-oriented 28 nm GaAs QWs (a) and (110)-oriented 28 nm GaAs QWs (c). Lower row: The black curves indicate the shape of $E_{\mathrm{THz}}$ after electro-optic detection, while the blue and red curves denote $\langle \dot{v} \rangle_a$ at early and late delay of the corresponding contour plots, respectively. The corresponding delay times are indicated in the contour plots with horizontal arrows. (e) to (h) show simulations of the experimental results plotted in (a) to (d), respectively.

# Detection of the anomalous velocity with sub-picosecond time resolution in semiconductor nanostructures

*Shekhar Priyadarshi,[*] Klaus Pierz, and Mark Bieler[†]*

Physikalisch-Technische Bundesanstalt, 38116 Braunschweig, Germany

## Supplementary Information

In this supplementary information we present a microscopic model which considers intrinsic and extrinsic effects in order to describe the temporal evolution of the anomalous velocity $\langle v \rangle_a$. The fact that we are only interested in the temporal evolution of $\langle v \rangle_a$ but not in its strength greatly simplifies our model. This is because we do not have to integrate the microscopic velocities over momentum, but can use simple symmetry relations to obtain the macroscopic temporal responses. In section I, we introduce the velocity matrix leading to the different components of $\langle v \rangle_a$ [1]. In section II, we introduce the perturbation approach allowing us to obtain time-domain responses of intrinsic (section III) and extrinsic (section IV) components of $\langle v \rangle_a$. Finally, we model the emitted terahertz pulses $\langle \dot{v} \rangle_a$ in section V.

### I. Anomalous velocity

For the calculation of observable macroscopic velocities in semiconductors we use the density matrix approach, where the macroscopic velocity is given by $\langle \mathbf{v} \rangle = \overline{Tr}(\mathbf{v}\rho)$, with $\rho$ denoting the density matrix, and $\overline{Tr}$ expressing the momentum-space average of the trace. The microscopic velocity matrix $\mathbf{v}$ is defined as [1,2]:

$$\mathbf{v} = \left( \mathbf{v}_{nn} + \mathbf{v}_{sj,nn} - \frac{e}{\hbar} \mathbf{E}_{THz} \times \mathbf{\Omega}_{nn} \right). \tag{1}$$

Here, $\mathbf{v}_{nn}$ is the normal velocity given by the energy dispersion and n is the band index. Intraband motions are manipulated by the electric-terahertz field ($\mathbf{E}_{THz}$) which along with the Berry curvature ($\mathbf{\Omega}_{nn} = \nabla_\mathbf{k} \times \mathbf{\xi}_{nn}$) contributes to the intrinsic part of the anomalous velocity. $\mathbf{\xi}_{nn} = i\langle u_\mathbf{k} | \nabla_\mathbf{k} | u_\mathbf{k} \rangle$ is the Berry connection as obtained from the Bloch wavefunctions $u_\mathbf{k}$ [1].

The component $\mathbf{v}_{sj,nn}$, is commonly known as side-jump velocity and is one of the origins of extrinsic effects. It results from carrier-impurity [2] or carrier-phonon [3] scattering and is given by $\mathbf{v}_{sj,nn} = \sum_{\mathbf{k}'} w_{\mathbf{k}'\mathbf{k}} \delta \mathbf{r}_{\mathbf{k}'\mathbf{k},nn}$, where $w_{\mathbf{k}'\mathbf{k}}$ is the scattering probability and $\delta \mathbf{r}_{\mathbf{k}'\mathbf{k},nn} = \mathbf{\xi}_{nn}(\mathbf{k}') - \mathbf{\xi}_{nn}(\mathbf{k}) - (\nabla_\mathbf{k} + \nabla_{\mathbf{k}'}) \arg(\langle u_{\mathbf{k}'} | u_\mathbf{k} \rangle)$ is the coordinate shift of carriers during scattering with impurities or phonons [2,3]. The subscripts $\mathbf{k}'\mathbf{k}$ indicate that the scattering event changes the carrier momentum from $\mathbf{k}$ to $\mathbf{k}'$. So far, contributions from phonons have not been considered for the theoretical description of extrinsic anomalous velocities. Yet there is no reason why phonons should not contribute to extrinsic effects [3].

For the calculation of $\langle \mathbf{v} \rangle_a$ we are only interested in the velocity perpendicular to the electric terahertz field ($\mathbf{E}_{THz} = E_{THz} \hat{x}$). Therefore, we evaluate $\langle \mathbf{v} \rangle$ in y-direction, in the following referred to as $\langle v \rangle_a$ [1,2]:

$$\langle v \rangle_a = \int \frac{d\mathbf{k}}{4\pi^2} \left\{ \sum_n \left( v_{nn}^y + v_{sj}^y \right) \rho_{nn,\mathbf{k}} + \frac{e}{\hbar} E_{THz} \sum_n \Omega_{nn}^z \rho_{nn,\mathbf{k}} \right\}. \tag{2}$$

---

[*] e-mail: shekhar.priyadarshi@ptb.de
[†] e-mail: mark.bieler@ptb.de

## II. Perturbation approach

The diagonal elements of the density matrix will be obtained from a perturbation approach. In our experiment, we optically excite heavy-hole excitons in GaAs quantum-well samples. To model such excitations, we consider a four-level system constituting two conduction bands ($|\bar{2}\rangle = |-1/2\rangle$ and $|2\rangle = |+1/2\rangle$) and two heavy-hole bands ($|\bar{1}\rangle = |-3/2\rangle$ and $|1\rangle = |+3/2\rangle$). Bars on the band indices indicate spin-down subbands. The allowed spin-up and spin-down transitions $|1\rangle \leftrightarrow |2\rangle$ and $|\bar{1}\rangle \leftrightarrow |\bar{2}\rangle$, respectively, result in excitation of excitons and free carriers. In the calculations, we start with two-level density-matrix equations in the length gauge for the spin-up transitions and later on use symmetry relations to account for spin-down transitions [4–7]:

$$\dot{\rho}_{12,\mathbf{k}} = -(\gamma + i\omega_{12,\mathbf{k}})\rho_{12,\mathbf{k}} - i\frac{e}{\hbar}\mathbf{r}_{12}\cdot\mathbf{E}(2\rho_{11,\mathbf{k}} - 1) - \frac{e}{\hbar}\{\nabla_{\mathbf{k}}\rho_{12,\mathbf{k}} - i(\boldsymbol{\xi}_{11} - \boldsymbol{\xi}_{22})\rho_{12,\mathbf{k}}\}\cdot\hat{x}E_{\text{THz}}, \quad (3)$$

$$\dot{\rho}_{11,\mathbf{k}} = -i\frac{e}{\hbar}(\rho_{12,\mathbf{k}}\mathbf{r}_{21}\cdot\mathbf{E} - \rho_{21,\mathbf{k}}\mathbf{r}_{12}\cdot\mathbf{E}) - \frac{e}{\hbar}\nabla_{\mathbf{k}}\rho_{11,\mathbf{k}}\cdot\hat{x}E_{\text{THz}} - \dot{\rho}_{11,\mathbf{k},\text{coll}}. \quad (4)$$

Here, $\mathbf{r}_{12}$ and $\mathbf{r}_{21}$ are the transition dipole matrix elements with $\mathbf{r}_{21}(\mathbf{k}) = \mathbf{r}_{\overline{12}}(-\mathbf{k})$ [8]; the electric field of the optical pulse is given by $\mathbf{E}(t) = \mathbf{E}_{\text{env}}(t)\exp(-i\omega_c t) + c.c.$, where $\mathbf{E}_{\text{env}}(t)$ is the electric-field envelope and $\omega_c$ is the carrier frequency. Moreover, $\omega_{12}$ is the resonance frequency of the two-level system, $\gamma$ is the dephasing rate, and $\dot{\rho}_{11,\mathbf{k},\text{coll}}$ is the collision term. In equations (3) and (4) we assumed that $\mathbf{E}$ excites the population and polarization while $E_{\text{THz}}$ accelerates the population and polarization.

We will solve equations (3) and (4) using the perturbation approach with $\mathbf{E}$ and $E_{\text{THz}}$ being the perturbation. The lowest-order populations relevant to us are $\rho_{11,\mathbf{k}}^{(2)}$ and $\rho_{11,\mathbf{k}}^{(3)}$. While $\rho_{11,\mathbf{k}}^{(2)}$ is second order in $\mathbf{E}$, $\rho_{11,\mathbf{k}}^{(3)}$ is obtained from $\rho_{11,\mathbf{k}}^{(2)}$ with $E_{\text{THz}}$ as perturbation. With these perturbation solutions and the symmetry relations $\mathbf{v}_{nn}(-\mathbf{k}) = -\mathbf{v}_{\bar{n}\bar{n}}(\mathbf{k})$ [8], $\boldsymbol{\Omega}_{nn}(\mathbf{k}) = -\boldsymbol{\Omega}_{\bar{n}\bar{n}}(-\mathbf{k})$ [9], and $\mathbf{v}_{\text{sj},\bar{n}\bar{n}}(-\mathbf{k}) = \mathbf{v}_{\text{sj},nn}(\mathbf{k})$ [3,10] we can write equation (2) as two separate equations, both of which account for both allowed transitions $|1\rangle \leftrightarrow |2\rangle$ and $|\bar{1}\rangle \leftrightarrow |\bar{2}\rangle$:

$$\langle v\rangle_{\text{a,int}} = \frac{e}{\hbar}E_{\text{THz}}\int\frac{d\mathbf{k}}{4\pi^2}\Omega_{11}^z(\mathbf{k})\{\rho_{11,\mathbf{k}}^{(2)} - \rho_{\bar{1}\bar{1},-\mathbf{k}}^{(2)}\}, \quad (5)$$

$$\langle v\rangle_{\text{a,ext}} = \int\frac{d\mathbf{k}}{4\pi^2}v_{11}^y\{\rho_{11,\mathbf{k}}^{(3)} - \rho_{\bar{1}\bar{1},-\mathbf{k}}^{(3)}\} + \int\frac{d\mathbf{k}}{4\pi^2}v_{\text{sj},11}^y\{\rho_{11,\mathbf{k}}^{(3)} + \rho_{\bar{1}\bar{1},-\mathbf{k}}^{(3)}\}. \quad (6)$$

Here, $\langle v\rangle_{\text{a,int}}$ and $\langle v\rangle_{\text{a,ext}}$ express the intrinsic and extrinsic parts, respectively, of the anomalous velocity. We consider only heavy-hole bands in the calculation of $\langle v\rangle_a$ since in the experiments we resonantly excited heavy-hole excitons with negligible contributions from light-hole transitions. As will be seen later $\langle v\rangle_{\text{a,ext}}$ includes contributions from anomalous distributions and asymmetric scattering of carriers, as well as from the side-jump velocity.

## III. Intrinsic $\langle v\rangle_a$

For the calculation of the intrinsic component we have to obtain an expression for the second-order populations $\rho_{11,\mathbf{k}}^{(2)}$ and $\rho_{\bar{1}\bar{1},-\mathbf{k}}^{(2)}$. We assume the zero-order solution $\rho_{11}^{(0)}(t,\mathbf{k})$ to be zero, which is a valid approximation for optically-excited undoped semiconductors. Next, we set $\mathbf{E}_{\text{env}}(t) = \hat{x}E_{\text{env}}(t)i + \hat{y}E_{\text{env}}(t)$ for circularly-polarized optical pulses, take $\omega_{12} = \omega_c$ for resonant excitation, and calculate the polarization being first order in $\mathbf{E}(t)$:

$$\rho_{12,\mathbf{k}}^{(1)}(t) = i\,{}^e\!/_\hbar \{i\, r_{12}^x + r_{12}^y\} f(t)\exp(-i\omega_c t)\,, \tag{7}$$

with, $f(t) = E_{\text{env}}(t) * [u(t)\exp(-\gamma t)]$, $u(t)$ being the Heaviside step function, and $*$ denoting convolution.

We then calculate $\rho_{11,\mathbf{k}}^{(2)}$ from equations (7) and (4). This gives:

$$\dot{\rho}_{11,\mathbf{k}}^{(2)}(t) = g_{11,\mathbf{k}} I(t) - \dot{\rho}_{11,\mathbf{k},\text{coll}}(t)\,. \tag{8}$$

Here, $g_{11,\mathbf{k}} = -\frac{2e^2}{\hbar^2} Im\{r_{12}^x r_{21}^y - r_{12}^y r_{21}^x\} = -g_{\bar{1}\bar{1},-\mathbf{k}}$ is the initial momentum-space carrier distribution (distribution without any scattering) and $I(t) = f(t) E_{\text{env}}(t)$. It is important to note that $\rho_{11,\mathbf{k}}^{(2)}$ in equation (8) represents the component of the population which is dependent on the helicity of the circularly-polarized excitation.

To solve equation (8) we adopt the relaxation time approximation for the collision term $\dot{\rho}_{11,\mathbf{k},\text{coll}}(t) = w^{\text{m}}\left(\rho_{11,\mathbf{k}}^{(2)}(t) - \bar{\rho}_{11,\mathbf{k}}^{(2)}(t)\right)$, with $w^{\text{m}}$ being the rate of relaxation in momentum space which includes both momentum and energy relaxation. Here we assume that the energy and momentum relaxation in the valence band are correlated. (Phonon scattering is dominant for both relaxation processes leading to similar relaxation rates.) The bar over population terms denotes the relaxed population. This term does not change with momentum and energy relaxation, but with spin relaxation. With above definitions we can also write $\bar{g}_{11,\mathbf{k}} = -\bar{g}_{\bar{1}\bar{1},-\mathbf{k}}$ and $\bar{\rho}_{11,\mathbf{k}}^{(2)}(t) = -\bar{\rho}_{\bar{1}\bar{1},-\mathbf{k}}^{(2)}(t)$. The difference $\bar{\rho}_{11,\mathbf{k}}^{(2)}(t) - \bar{\rho}_{\bar{1}\bar{1},-\mathbf{k}}^{(2)}(t)$, which gives the spin-polarization, decays with spin relaxation $w^{\text{sf}}$ and is then obtained from:

$$\dot{\bar{\rho}}_{11,\mathbf{k}}^{(2)}(t) - \dot{\bar{\rho}}_{\bar{1}\bar{1},-\mathbf{k}}^{(2)}(t) = 2\bar{g}_{11,\mathbf{k}} I(t) - w^{\text{sf}}\left(\bar{\rho}_{11,\mathbf{k}}^{(2)}(t) - \bar{\rho}_{\bar{1}\bar{1},-\mathbf{k}}^{(2)}(t)\right). \tag{9}$$

Inserting the solution of equation (9) into equation (8) and assuming that $w^{\text{m}} \gg w^{\text{sf}}$ yields:

$$\rho_{11,\mathbf{k}}^{(2)}(t) = \left[(g_{11,\mathbf{k}} - \bar{g}_{11,\mathbf{k}})f_{\text{m}}(t) + \bar{g}_{11,\mathbf{k}} f_{\text{s}}(t)\right] = -\rho_{\bar{1}\bar{1},-\mathbf{k}}^{(2)}(t)\,, \tag{10}$$

with $f_{\text{m}}(t) = I(t) * u(t)\exp(-w^{\text{m}} t)$ and $f_{\text{s}}(t) = I(t) * u(t)\exp(-w^{\text{sf}} t)$. The term $g_{11,\mathbf{k}}$ in equation (10) can be written as $g_{11,\mathbf{k}} = g_{11,\mathbf{k}}^o + g_{11,\mathbf{k}}^e$ with $g_{11,\mathbf{k}}^o = -g_{11,-\mathbf{k}}^o$ being an odd function of $\mathbf{k}$ (odd distribution) and $g_{11,\mathbf{k}}^e = g_{11,-\mathbf{k}}^e$ being an even function of $\mathbf{k}$ (even distribution). With this definition we rewrite equation (10):

$$\rho_{11,\mathbf{k}}^{(2)}(t) = \left[g_{11,\mathbf{k}}^o f_{\text{m}}(t) + (g_{11,\mathbf{k}}^e - \bar{g}_{11,\mathbf{k}})f_{\text{m}}(t) + \bar{g}_{11,\mathbf{k}} f_{\text{s}}(t)\right] = -\rho_{\bar{1}\bar{1},-\mathbf{k}}^{(2)}(t). \tag{11}$$

The first term in equation (11) describes momentum relaxation of an odd distribution, the second term describes momentum relaxation of an even distribution ($\bar{g}_{11,\mathbf{k}}$ is also an even function of $\mathbf{k}$), and the last term describes spin relaxation of an even distribution. Inserting equation (11) into equation (5) we then obtain an expression for $\langle v \rangle_{\text{a,int}}$:

$$\langle v \rangle_{\text{a,int}} = \frac{2e}{\hbar} E_{\text{THz}}(t) \int \frac{d\mathbf{k}}{4\pi^2} \Omega_{11}^z(\mathbf{k})\left[(g_{11,\mathbf{k}}^e - \bar{g}_{11,\mathbf{k}})f_{\text{m}}(t) + \bar{g}_{11,\mathbf{k}} f_{\text{s}}(t)\right], \tag{12}$$

where the term containing $g_{11,\mathbf{k}}^o$ cancelled since we assume $\mathbf{\Omega}_{11}(\mathbf{k})$ to be an even function of $\mathbf{k}$. In general $\langle v \rangle_{\text{a,int}}$ will be influenced by both energy and spin relaxation. However, if $\Omega_{11}^z(\mathbf{k})$ is zero at $\mathbf{k} = 0$, which is the case for the first heavy hole band in GaAs quantum wells [11,12], the terms containing $\bar{g}_{11,\mathbf{k}}$ will vanish such that $\langle v \rangle_{a,\text{int}}$ only decays with energy relaxation:

$$\langle v \rangle_{a,\text{int}} = \frac{2e}{\hbar} E_{\text{THz}}(t) \int \frac{d\mathbf{k}}{4\pi^2} \Omega_{11}^z(\mathbf{k}) g_{11,\mathbf{k}}^e f_m(t) . \tag{13}$$

**IV. Extrinsic $\langle v \rangle_a$**

The third-order population $\rho_{11,\mathbf{k}}^{(3)}$ will be obtained by solving [2]:

$$\dot{\rho}_{11,\mathbf{k}}^{(3)} = -\frac{e}{\hbar} \nabla_{\mathbf{k}} \rho_{11,\mathbf{k}}^{(2)} \cdot \hat{x} E_{\text{THz}} - \sum_{\mathbf{k}'} \left\{ w_{\mathbf{k}'\mathbf{k}} \rho_{11,\mathbf{k}}^{(3)} - w_{\mathbf{k}\mathbf{k}'} \rho_{11,\mathbf{k}'}^{(3)} - e w_{\mathbf{k}'\mathbf{k}}^s \frac{d\rho_{11,\mathbf{k}}^{(2)}}{d\epsilon_{\mathbf{k}}} E_{\text{THz}} \cdot \delta \mathbf{r}_{\mathbf{k}'\mathbf{k}} \right\} . \tag{14}$$

The term $\sum_{\mathbf{k}'} e \left\{ w_{\mathbf{k}'\mathbf{k}}^s \frac{d\rho_{11,\mathbf{k}}^{(2)}}{d\epsilon_{\mathbf{k}}} E_{\text{THz}} \cdot \delta \mathbf{r}_{\mathbf{k}'\mathbf{k}} \right\} = \frac{e}{\hbar} (\nabla_{\mathbf{k}} \rho_{11,\mathbf{k}}^{(2)} \cdot \mathbf{v}_{11}^i) E_{\text{THz}} \hat{x} \cdot \mathbf{v}_{\text{sj},11}$ in equation (14), with $\mathbf{v}_{11}^i = \hat{x}/v_{11}^x + \hat{y}/v_{11}^y$, accounts for carriers losing or gaining energy during scattering from $\mathbf{k}$ to $\mathbf{k}'$ [2]. The factor $w_{\mathbf{k}'\mathbf{k}}^s$ denotes a symmetric scattering probability being symmetric under the exchange of $\mathbf{k}$ and $\mathbf{k}'$ and $\epsilon_{\mathbf{k}}$ is the energy dispersion. For small momentum change $\boldsymbol{\kappa}$ of $\rho_{11,\mathbf{k}}^{(3)}$ as compared to $\rho_{11,\mathbf{k}}^{(2)}$ we can write:

$$\rho_{11,\mathbf{k}}^{(3)} \sim \nabla_{\mathbf{k}} \rho_{11,\mathbf{k}}^{(2)} \cdot \boldsymbol{\kappa} , \tag{15}$$

$$\dot{\rho}_{11,\mathbf{k}}^{(3)} \sim \nabla_{\mathbf{k}} \rho_{11,\mathbf{k}}^{(2)} \cdot \frac{d\boldsymbol{\kappa}}{dt} + \nabla_{\mathbf{k}} \dot{\rho}_{11,\mathbf{k}}^{(2)} \cdot \boldsymbol{\kappa} . \tag{16}$$

Moreover we assume small scattering angles, i.e., $\rho_{11,\mathbf{k}}^{(2)} \sim \rho_{11,\mathbf{k}'}^{(2)}$ and express the scattering term $w_{\mathbf{k}\mathbf{k}'}$ using the prescription of Ref. [10]:

$$w_{\mathbf{k}\mathbf{k}'} = w_{\mathbf{k}\mathbf{k}'}^s + \sigma w_{\mathbf{k}\mathbf{k}'}^a (\hat{k} \times \hat{k}')_z . \tag{17}$$

Here, $w_{\mathbf{k}\mathbf{k}'}^a$ denotes the anti-symmetric part of the scattering probability. It is important to emphasize that $w_{\mathbf{k}\mathbf{k}'}^a$ is symmetric under the exchange of $\mathbf{k}$ and $\mathbf{k}'$. The anti-symmetry is given by the factor $(\hat{k} \times \hat{k}')_z$ in equation (17). Moreover, $\sigma = \pm 1$, depending on the spin. With the help of equation (17) and for $\sigma = 1$ we can rewrite the first two scattering terms of equation (14) as follows:

$$\sum_{\mathbf{k}'} \left\{ w_{\mathbf{k}'\mathbf{k}} \rho_{11,\mathbf{k}}^{(3)} - w_{\mathbf{k}\mathbf{k}'} \rho_{11,\mathbf{k}'}^{(3)} \right\} = \nabla_{\mathbf{k}} \rho_{11,\mathbf{k}}^{(2)} \cdot \sum_{\mathbf{k}'} \left\{ w_{\mathbf{k}'\mathbf{k}} \boldsymbol{\kappa} - w_{\mathbf{k}\mathbf{k}'} \boldsymbol{\kappa}' \right\} = \nabla_{\mathbf{k}} \rho_{11,\mathbf{k}}^{(2)} \cdot (\boldsymbol{\kappa} w_{\mathbf{k}}^s + \boldsymbol{\kappa}_\perp w_{\mathbf{k}}^a) . \tag{18}$$

with $\boldsymbol{\kappa}_\perp = -\kappa^y \hat{x} + \kappa^x \hat{y}$ and $w_{\mathbf{k}}^s$ and $w_{\mathbf{k}}^a$ being symmetric and antisymmetric scattering rates respectively. The above definitions allow us to simplify equation (14):

$$\nabla_{\mathbf{k}} \rho_{11,\mathbf{k}}^{(2)} \cdot \left[ \dot{\boldsymbol{\kappa}} + \frac{e}{\hbar} \{ \hat{x} - v_{\text{sj},11}^x \mathbf{v}_{11}^i \} E_{\text{THz}} + (w_{\mathbf{k}}^s \boldsymbol{\kappa} \mp w_{\mathbf{k}}^a \boldsymbol{\kappa}_\perp) \right] + \nabla_{\mathbf{k}} \dot{\rho}_{11,\mathbf{k}}^{(2)} \cdot \boldsymbol{\kappa} = 0 . \tag{19}$$

We ignore the last term in equation (19) since it is temporally limited to the optical pulse which is very short in time (150 fs). This allows us to obtain a solution for $\boldsymbol{\kappa}$ and with equation (14) also a solution for the third-order population:

$$\rho_{11,\mathbf{k}}^{(3)}(t) = \left[ \sigma_x \frac{\partial \rho_{11,\mathbf{k}}^{(2)}(t)}{\partial k^x} + \sigma_y \frac{\partial \rho_{11,\mathbf{k}}^{(2)}(t)}{\partial k^y} \right] b(t) + \left[ \sigma_y \frac{\partial \rho_{11,\mathbf{k}}^{(2)}(t)}{\partial k^x} - \sigma_x \frac{\partial \rho_{11,\mathbf{k}}^{(2)}(t)}{\partial k^y} \right] w_{\mathbf{k}}^a a(t) , \tag{20}$$

$$\rho_{\overline{11},-\mathbf{k}}^{(3)}(t) = -\left[\sigma_x' \frac{\partial \rho_{11,\mathbf{k}}^{(2)}(t)}{\partial k^x} + \sigma_y \frac{\partial \rho_{11,\mathbf{k}}^{(2)}(t)}{\partial k^y}\right] b(t) + \left[\sigma_y \frac{\partial \rho_{11,\mathbf{k}}^{(2)}(t)}{\partial k^x} - \sigma_x' \frac{\partial \rho_{11,\mathbf{k}}^{(2)}(t)}{\partial k^y}\right] w_{\mathbf{k}}^{\text{a}} a(t). \qquad (21)$$

Here, $\sigma_x = \frac{e}{\hbar}\left(\frac{v_{\text{sj},11}^x}{v_{11}^x} - 1\right)$, $\sigma_x' = \frac{e}{\hbar}\left(\frac{v_{\text{sj},11}^x}{v_{11}^x} + 1\right)$, and $\sigma_y = \frac{e}{\hbar}\left(\frac{v_{\text{sj},11}^x}{v_{11}^y}\right)$. For $w_{\mathbf{k}}^{\text{a}} \ll w_{\mathbf{k}}^{\text{s}}$ the time-dependent functions $a(t)$ and $b(t)$ simplify to:

$$a(t) \sim E_{\text{THz}}(t) * [u(t)\, t \exp(-w_{\mathbf{k}}^{\text{s}} t)], \qquad (22)$$

$$b(t) \sim E_{\text{THz}}(t) * [u(t) \exp(-w_{\mathbf{k}}^{\text{s}} t)]. \qquad (23)$$

With equations (6) and (11) we then obtain three components for $\langle v \rangle_{\text{a,ext}}$:

$$\langle v \rangle_{\text{a,ext,adist}} = -4e \int \frac{d\mathbf{k}}{4\pi^2} v_{11}^y v_{\text{sj}}^x \left\{ \frac{d\bar{g}_{11,\mathbf{k}}}{d\epsilon_{\mathbf{k}}} f_s(t) + \frac{d(g_{11,\mathbf{k}}^e - \bar{g}_{11,\mathbf{k}})}{d\epsilon_{\mathbf{k}}} f_m(t) \right\} b(t), \qquad (24)$$

$$\langle v \rangle_{\text{a,ext,sj}} = 2e \int \frac{d\mathbf{k}}{4\pi^2} v_{\text{sj}}^y v_{11}^x \left\{ \frac{d\bar{g}_{11,\mathbf{k}}}{d\epsilon_{\mathbf{k}}} f_s(t) + \frac{d(g_{11,\mathbf{k}}^e - \bar{g}_{11,\mathbf{k}})}{d\epsilon_{\mathbf{k}}} f_m(t) \right\} b(t), \qquad (25)$$

$$\langle v \rangle_{\text{a,ext,sk}} = -2e \int \frac{d\mathbf{k}}{4\pi^2} (v_{11}^y)^2 \left\{ \frac{d\bar{g}_{11,\mathbf{k}}}{d\epsilon_{\mathbf{k}}} f_s(t) + \frac{d(g_{11,\mathbf{k}}^e - \bar{g}_{11,\mathbf{k}})}{d\epsilon_{\mathbf{k}}} f_m(t) \right\} w_{\mathbf{k}}^{\text{a}} a(t). \qquad (26)$$

Here, $\langle v \rangle_{\text{a,ext,adist}}$ results from the anomalous distribution of carriers due to coordinate shifts in the external electric field, $\langle v \rangle_{\text{a,ext,sj}}$ results from the side-jump velocity [2]. Since the temporal dynamics of $\langle v \rangle_{\text{a,ext,adist}}$ and $\langle v \rangle_{\text{a,ext,sj}}$ are same and their microscopic origins involve coordinate shift of carriers, we do not distinguish between them further and only consider the side-jump contribution in the following. Finally, $\langle v \rangle_{\text{a,ext,sk}}$ is the skew-scattering contribution to $\langle v \rangle_{\text{a}}$ resulting from anti-symmetric scattering.

In equations (24) to (26) we have already replaced $\rho_{11,\mathbf{k}}^{(2)}$ with the expression given by equation (11) and considered that $v_{11}$ and $v_{\text{sj}}$ are odd functions of $\mathbf{k}$ (see figure 1 of the main paper). This eliminates the contribution in equation (11) which is an odd function of $\mathbf{k}$ (influenced by momentum relaxation) and only leaves the even-in-$\mathbf{k}$ contribution being influenced by energy and spin relaxations. It should be emphasized that the results presented in equations (12), (13), and (24) to (26) also remain valid in case of spin splitting being linear in $\mathbf{k}$. (For such situations we simply perform a $\mathbf{k}$-space transformation $(\mathbf{k} - \mathbf{k}_0) \to \mathbf{k}$, with $\mathbf{k}_0$ being the band extremum and repeat the above analysis.)

## V. THz emission from $\langle v \rangle_{\text{a}}$

The time-domain expressions for the intrinsic and extrinsic components of $\langle v \rangle_{\text{a}}$ resulting from the previous sections are identical with the simple intuitive explanation given in the main text. Yet, to allow for direct comparison to experiments, the terahertz radiation $\langle \dot{v} \rangle_{\text{a}}$ emitted from $\langle v \rangle_{\text{a}}$ has to be modeled. For this purpose we first introduce the time delay $\tau$ between the optical pulse and $E_{\text{THz}}$ in equations (13), (25), and (26), time differentiate the expressions, and convolve the results with $H_{\text{EOS}}(t)$ accounting for the transfer function of the electro-optic detection. $E_{\text{THz}}$ results from optical rectification in a 2 mm thick ZnTe crystal, while $H_{\text{EOS}}$ results from electro-optic detection in a 1 mm thick ZnTe crystal. For both time-domain functions, well established models exist [13–17].

With the above definition $\langle \dot{v} \rangle_{\text{a,int}}^y$, $\langle \dot{v} \rangle_{\text{a,ext,sj}}^y$, and $\langle \dot{v} \rangle_{\text{a,ext,sk}}^y$ become functions of $t$ and $\tau$. We visualize these functions as contour plots in the same manner as the experimental results.